\documentclass[a4paper,11pt]{article}
\usepackage{pos}

\usepackage{graphicx}
\usepackage{dcolumn}% Align table columns on decimal point
\usepackage{bm,graphicx,hyperref}
\usepackage{float,amsfonts,graphicx}
\usepackage{color}
\usepackage{tikz}
\usepackage{physics}
\usepackage{adjustbox}
\usetikzlibrary{decorations.pathreplacing}
\usetikzlibrary{arrows.meta}
\usepackage{bm,hyperref}
\relax

\usepackage{subcaption}

\title{Computing quantum entanglement with machine learning}
\author*[a]{Andrea Bulgarelli}
\author[b]{Elia Cellini}
\author[c,d]{Karl Jansen}
\author[d]{Stefan K\"{u}hn}
\author[e]{Alessandro Nada}
\author[f,g,h]{Shinichi Nakajima}
\author[a,i]{Kim A. Nicoli}
\author[e,l]{Marco Panero}

\affiliation[a]{Transdisciplinary Research Area ``Building Blocks of Matter and Fundamental Interactions'' (TRA Matter) and Helmholtz Institute for Radiation and Nuclear Physics (HISKP), University of Bonn, Nussallee 14-16, 53115 Bonn, Germany}
\affiliation[b]{Higgs Centre for Theoretical Physics, School of Physics and Astronomy,
The University of Edinburgh, Edinburgh EH9 3FD, United Kingdom}
\affiliation[c]{Computation-Based Science and Technology Research Center, The Cyprus Institute,  Nicosia, Cyprus}
\affiliation[d]{Deutsches Elektronen-Synchrotron DESY, Zeuthen, Germany}
\affiliation[e]{Physics Department, University of Turin \& INFN, Turin unit, via Pietro Giuria 1, I-10125 Turin, Italy}
\affiliation[f]{Berlin Institute for the Foundations of Learning and Data (BIFOLD), Berlin, Germany}
\affiliation[g]{Machine Learning Group, Technische Universit\"{a}t Berlin, Berlin, Germany}
\affiliation[h]{RIKEN Center for AIP, Tokyo, Japan}
\affiliation[i]{Oldendorff Carriers GmbH \& Co. KG, L\"{u}beck, Germany}
\affiliation[l]{Department of Physics and Helsinki Institute of Physics, PL 64, FIN-00014 University of Helsinki, Finland}

\emailAdd{abulgare@uni-bonn.de}

\abstract{Entanglement calculations in quantum field theories are extremely challenging and typically rely on the replica trick, where the problem is rephrased in a study of defects. We demonstrate that the use of deep generative models drastically outperforms standard Monte Carlo algorithms. Remarkably, such a machine-learning method enables high-precision estimates of Rényi entropies in three dimensions for very large lattices. Moreover, we propose a new paradigm for studying lattice defects with flow-based sampling.
\vspace{.5cm}
\begin{flushright}
HIP-2025-33/TH
\end{flushright}
}

\FullConference{The 42nd International Symposium on Lattice Field Theory (LATTICE2025)\\
2-8 November 2025\\
Tata Institute of Fundamental Research, Mumbai, India\\}

\begin{document}
\maketitle

\section{Introduction}

Following~\cite{Bulgarelli:2024yrz}, in this contribution we discuss a numerical algorithm, based on machine learning methods, to compute quantum entanglement in lattice field theories. This is motivated by the fact that quantum entanglement, while rooted in the subjects of quantum information theory and quantum mechanics, has applications that extend across various domains of physics. In quantum field theories and quantum many-body systems, non-local correlations provide access to information that is typically difficult to obtain through local probes. For instance, the central charge of $(1+1)$-dimensional conformal field theories can be determined by analyzing the scaling behavior of the entanglement entropy with the subsystem size~\cite{Calabrese:2004eu}. More generally, entanglement measures play a significant role in condensed matter physics~\cite{Vidal:2002rm}, quantum gravity and the gauge/gravity correspondence~\cite{Ryu:2006bv} and confinement~\cite{Klebanov:2007ws}, among others.

A widely used measure of entanglement is entropy. For a system described by a density matrix $\rho$, and given a bipartition of the Hilbert space $\mathcal{H} = \mathcal{H}_A \otimes \mathcal{H}_B$, the reduced density matrix is defined as
\begin{align}
\rho_A = \Tr_B \rho.
\end{align}
The Rényi entropies are then given by
\begin{align}
S_n = \frac{1}{1-n} \log \Tr \rho_A^n,
\end{align}
where $n\geq 2$ is an integer. It is well known that, in quantum field theories and in arbitrary spacetime dimension $D$, the Rényi entropies are ultraviolet-divergent quantities. For certain geometries, one can nevertheless isolate the physical, ultraviolet-finite contributions by considering the derivative of $S_n$ with respect to the linear size $l$ of the subsystem $A$,
\begin{align}
C_n = \frac{l^{D-1}}{|\partial A|} \frac{\partial S_n}{\partial l},
\label{entropic_c-function_definition}
\end{align}
where $|\partial A|$ denotes the area of the boundary of the subsystem. $C_n$, commonly referred to as the \textit{entropic c-function}, encodes universal information about the underlying quantum field theory while being well defined in the continuum.

Explicit calculations of such quantities present significant challenges, particularly in systems with many degrees of freedom (d.o.f.). While some of these difficulties are conceptual, especially in the context of gauge theories~\cite{Buividovich:2008gq, Casini:2013rba, Ghosh:2015iwa}, this contribution focuses on the computational challenges associated with estimating entanglement measures in generic, potentially strongly interacting quantum field theories. To this end, we consider systems defined on a lattice, and we work within the framework of the replica trick~\cite{Calabrese:2004eu}, which reduces the computation of entanglement monotones to evaluating ratios of partition functions.

Still, the challenge remains significant. Traditional Monte Carlo methods are not naturally suited for computing partition function ratios, as such ratios cannot generally be evaluated as primary observables. This limitation motivates the search for alternative approaches to compute entanglement-related quantities within the path-integral formalism. In recent years, flow-based sampling and generative models have emerged as a powerful numerical strategy both for sampling probability distributions~\cite{Albergo:2019eim, Nicoli:2019gun, Kanwar:2020xzo, Bacchio:2022vje, Cranmer:2023xbe, Caselle:2023mvh, Albandea:2023wgd, Caselle:2024ent, Singha:2025lsd, Kreit:2025cos, Schuh:2025gks, Vega:2025hgz} and for estimating quantities associated with ratios of partition functions~\cite{Caselle:2016wsw, Alba:2016bcp, Caselle:2018kap, Nicoli:2020njz, Bulgarelli:2023ofi, Bulgarelli:2024onj, Bialas:2024gha, Bulgarelli:2025riv}. Furthermore, connections with non-equilibrium Monte Carlo simulations~\cite{Caselle:2022acb} have enabled the development of flow-based samplers with fully controlled scaling properties~\cite{Bonanno:2024udh, Bulgarelli:2024brv, Bonanno:2025pdp}.

The purpose of our work~\cite{Bulgarelli:2024yrz} is to examine whether these techniques and, in particular, a specific instance of flow-based sampling, Normalizing Flows (NFs)~\cite{Rezende:2015}, are suitable for efficient and scalable computations of entropic c-functions.

\section{The replica trick}

\begin{figure}[t]
\begin{subfigure}{0.45\textwidth}
    \centering
    \includegraphics[width=\textwidth]{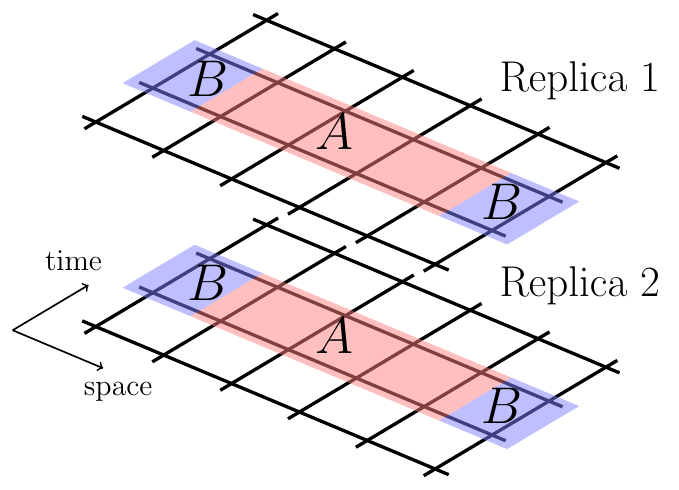}
\end{subfigure}
\begin{subfigure}{0.45\textwidth}
    \centering
    \includegraphics[width=\textwidth]{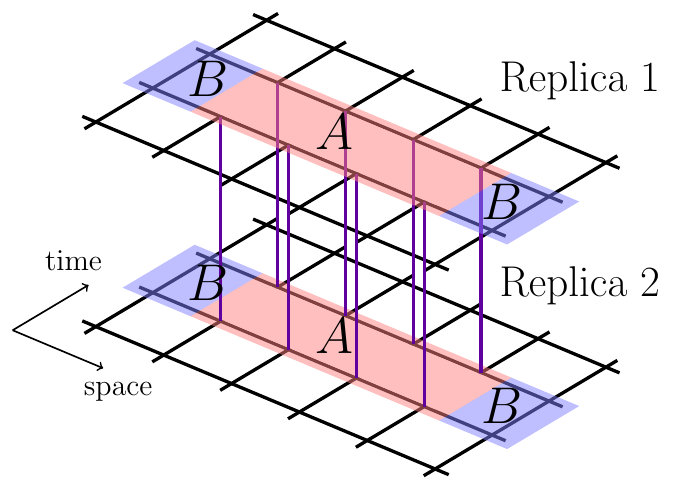}
\end{subfigure}
\caption{Left panel: system of two independent lattices. Right panel: the two replicas are coupled through a set of links along the subsystem $A$ at a fixed Euclidean time.}
\label{fig:replica_trick}
\end{figure}

The replica trick for calculating Rényi entropies was first introduced by Calabrese and Cardy in the continuum~\cite{Calabrese:2004eu}, and later adapted to lattice systems by Buividovich and Polikarpov~\cite{Buividovich:2008gq}. In this work, we focus directly on the lattice formulation\footnote{For a clear derivation in the continuum, see ref.~\cite{Casini:2009sr}.}. To compute the $n$-th Rényi entropy, one considers $n$ independent replicas of the same lattice model, so that the action of the replicated system is the sum of the actions of the individual copies. Denoting by $Z$ the partition function of the single theory, the replicated partition function is simply $Z^n$ (see fig.~\ref{fig:replica_trick}, left panel).

The subsystem $A$ is identified in each replica at a fixed Euclidean time. As illustrated in fig.~\ref{fig:replica_trick} (right panel), one then constructs a geometry of coupled replicas by introducing new links connecting different copies along subsystem $A$. Geometrically, this setup corresponds to a discretized Riemann surface, with $A$ lying along a branch cut. The partition function of the lattice system in this geometry is denoted by $Z_n$. One can then show~\cite{Calabrese:2004eu} that
\begin{align}
S_n = \frac{1}{1-n} \log \frac{Z_n}{Z^n}.
\label{replica_trick_result}
\end{align}
This relation enables the explicit computation of Rényi entropies as ratios of partition functions, a strategy used in Monte Carlo simulations~\cite{Buividovich:2008kq, Itou:2015cyu, Rabenstein:2018bri, Bulgarelli:2023ofi, Bulgarelli:2024onj}.

In this work, we are interested in the derivative of the Rényi entropy, which can also be accessed through the replica trick~\cite{Calabrese:2004eu}
\begin{align}
C_n = \frac{l^{D-1}}{|\partial A|}\frac{1}{1-n}\lim_{a\rightarrow 0}\frac{1}{a}\log \frac{Z_n(l+a)}{Z_n(l)},
\end{align}
where $a$ is the lattice spacing. Developing efficient algorithms to evaluate the ratio $Z_n(l+a)/Z_n(l)$ is the main goal of this study~\cite{Bulgarelli:2024yrz}.

\section{Flow-based sampling}

Traditional Monte Carlo simulations used to compute partition function ratios typically rely on combining results from independent runs. As a consequence, the ratio itself is not a primary observable directly accessible on the lattice. This limitation motivates the search for alternative approaches that enable a more direct estimation of entanglement-related quantities.

In recent years, flow-based sampling has emerged as a powerful tool for sampling probability distributions, particularly in systems affected by critical slowing down or severe ergodicity problems~\cite{Albergo:2019eim}. Unlike standard Markov Chain Monte Carlo methods, which sample the target probability distribution $p$ directly, flow-based approaches rely on two key components: sampling from a prior distribution $q$, and applying a transformation to the configurations sampled from $q$, which maps them to configurations sampled from $p$. As a byproduct of this procedure, one is able to directly measure the partition function ratio between the prior and the target distribution~\cite{Nicoli:2020njz}.

The specific instance of flow-based sampling considered in this work is that of NFs~\cite{Rezende:2015, Albergo:2019eim}. Normalizing flows are generative models constructed as compositions of diffeomorphisms between probability distributions. In general, the transformation can be expressed as
\begin{align}
g_{\theta}(\phi_0) = (g_N \circ \dots \circ g_1)(\phi_0),
\end{align}
where the individual transformations $g_i$ are referred to as coupling layers, and $\theta$ denotes the set of variational parameters. This mapping transforms the prior distribution $q$ into a variational distribution $q_{\theta}$, whose parameters are optimized to maximize the overlap with the target distribution $p$. A final Metropolis step or reweighting is applied to ensure exact asymptotic sampling from $p$.

A crucial feature of NFs is their ability to provide access to the unnormalized likelihood,
\begin{align}
q_{\theta}(g_\theta(\phi_0)) = q(\phi_0) J_{g_{\theta}}^{-1},
\end{align}
where $J_{g_{\theta}}$ is the Jacobian determinant of the transformation. This property allows one to compute the ratio of partition functions $Z_p / Z_q$~\cite{Nicoli:2020njz}. Defining the weights
\begin{align}
\tilde{w} = \exp[-(S_p(g_\theta(\phi_0)) - S_q(\phi_0) - \log J_{g_\theta})],
\end{align}
one obtains
\begin{align}
\frac{Z_p}{Z_q} = \langle \tilde{w} \rangle_{\phi_0 \sim q}.
\label{partition_function_ratio_from_NF}
\end{align}

Equation~\eqref{partition_function_ratio_from_NF} enables a direct calculation of the partition function ratio. However, explicit applications to large-scale lattice simulations are limited due to problematic scaling of such methods, which has been pointed out to be exponential with the number of d.o.f. in most of the cases~\cite{DelDebbio:2021qwf, Abbott:2022zsh, Abbott:2023thq}. In our work~\cite{Bulgarelli:2024yrz} we propose an approach to mitigate scaling limitations of NFs, and we apply it to large-scale studies of entropic c-functions.

\section{Defect coupling layers}

\begin{figure}[t]
\centering
\includegraphics[width=.5\textwidth]{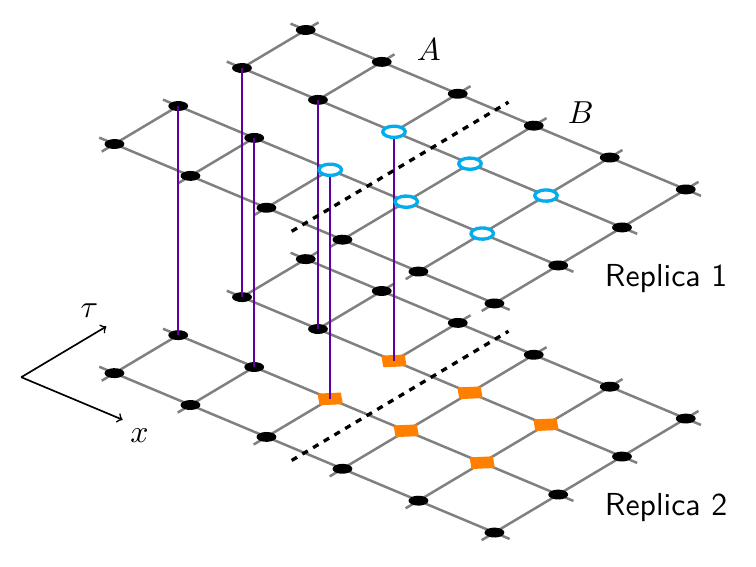}
\caption{$(1 + 1)$-dimensional replicated lattice ($\tau$ is the Euclidean-time direction). Purple links connect different replicas. The lattice is divided in three parts in the NF: the environment (black sites), which does not enter the coupling layer; frozen sites (empty cyan circles), that are inputs of the neural network; active sites (orange diamonds), which are transformed by the NF.}
\label{fig:replica_space_ml_defect}
\end{figure}

A crucial feature of the target partition-function ratio in eq.~\eqref{entropic_c-function_definition} is that the difference between the geometries defined by $Z_n(l)$ and $Z_n(l+a)$ can be traced back to a subset of lattice links located at the edge of the cut in replica space. We refer to this set of links as a ``defect''\footnote{In the continuum, the endpoint of the cut is, strictly speaking, a (possibly conformal) defect~\cite{Bianchi:2015liz}.}.

Our proposal is summarized in fig.~\ref{fig:replica_space_ml_defect}. For concreteness, in what follows we focus on a scalar $\phi^4$ field theory, where the lattice degrees of freedom are scalar fields defined on lattice sites, $\phi(x)$, and the action is
\begin{align}
S = \sum_x (1 - 2\lambda)\phi^2(x) + \lambda \phi^4(x) - 2\kappa \sum_{\mu=1}^D \phi(x)\phi(x + a\hat{\mu}).
\end{align}
First, configurations of the prior probability distribution (namely, the replica space with a cut of length $l$) are generated using standard Monte Carlo methods. A NF is then constructed to act only on a small subset of degrees of freedom near the endpoint of the branch cut. More precisely, we introduce the following transformation
\begin{align}
\phi'_{\rm active} = \exp(-|s(\phi_{\rm frozen})|)\,\phi_{\rm active} + t(\phi_{\rm frozen}),
\end{align}
which is referred to as a mask. A coupling layer is made of several masks that are combined together so that every degree of freedom (of the relevant region selected) is transformed exactly once. This is a standard construction ensuring the invertibility of the coupling layer~\cite{Dinh:2017}: the active fields (orange diamonds in fig.~\ref{fig:replica_space_ml_defect}) are transformed by the layer, while the frozen fields (cyan empty circles) serve as inputs to the neural network that outputs the functions $s$ and $t$. Both active and frozen sites are selected in a region localized around the defect, i.e. the endpoint of the cut, ensuring that only a small portion of the lattice enters the transformation. This strategy identifies a priori the physically relevant region on which the NF acts, thereby drastically reducing the number of degrees of freedom processed by the model.

\section{Numerical results}

\begin{figure}[t]
\centerline{\includegraphics[height=0.35\textwidth]{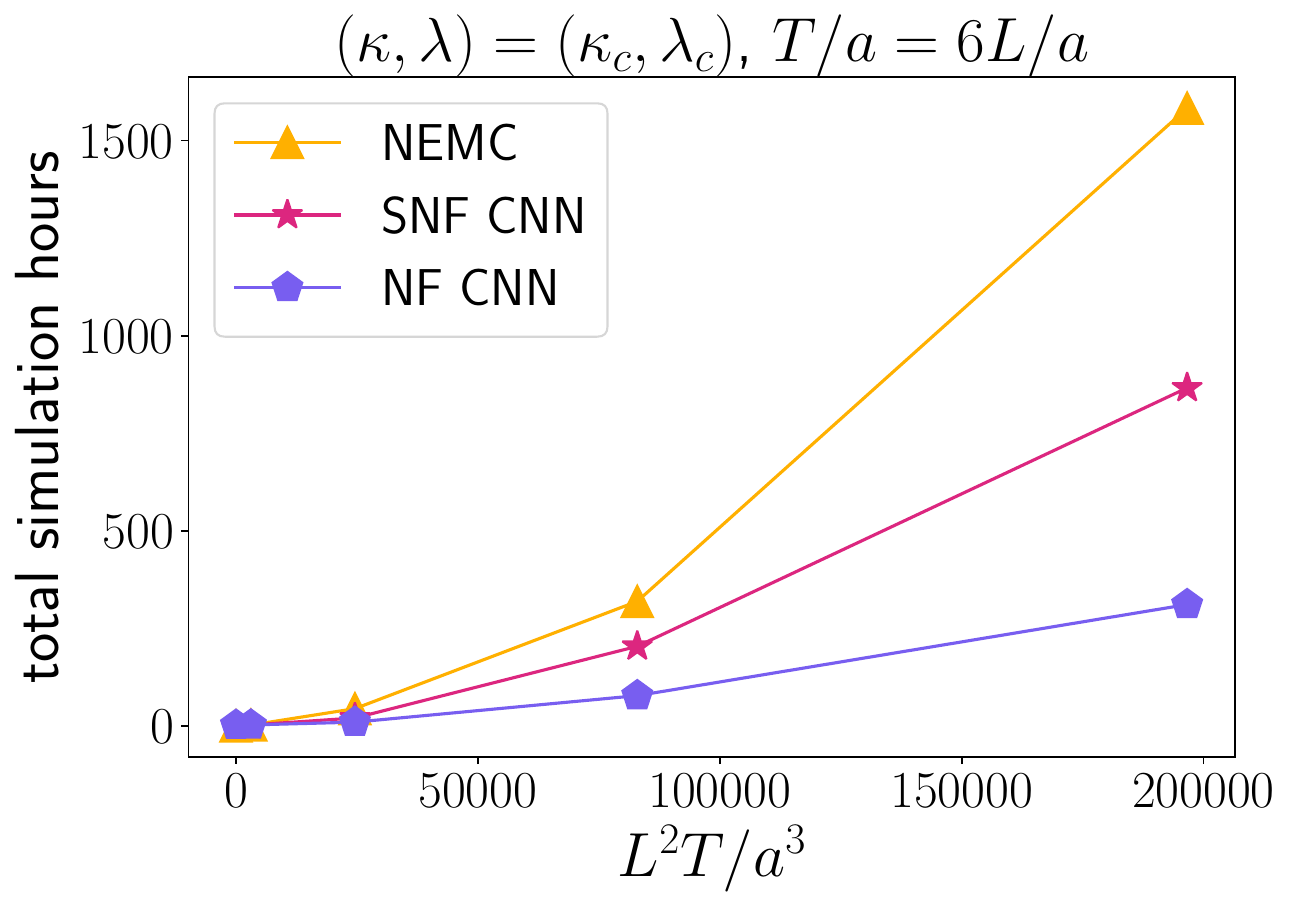} \hfill \includegraphics[height=0.35\textwidth]{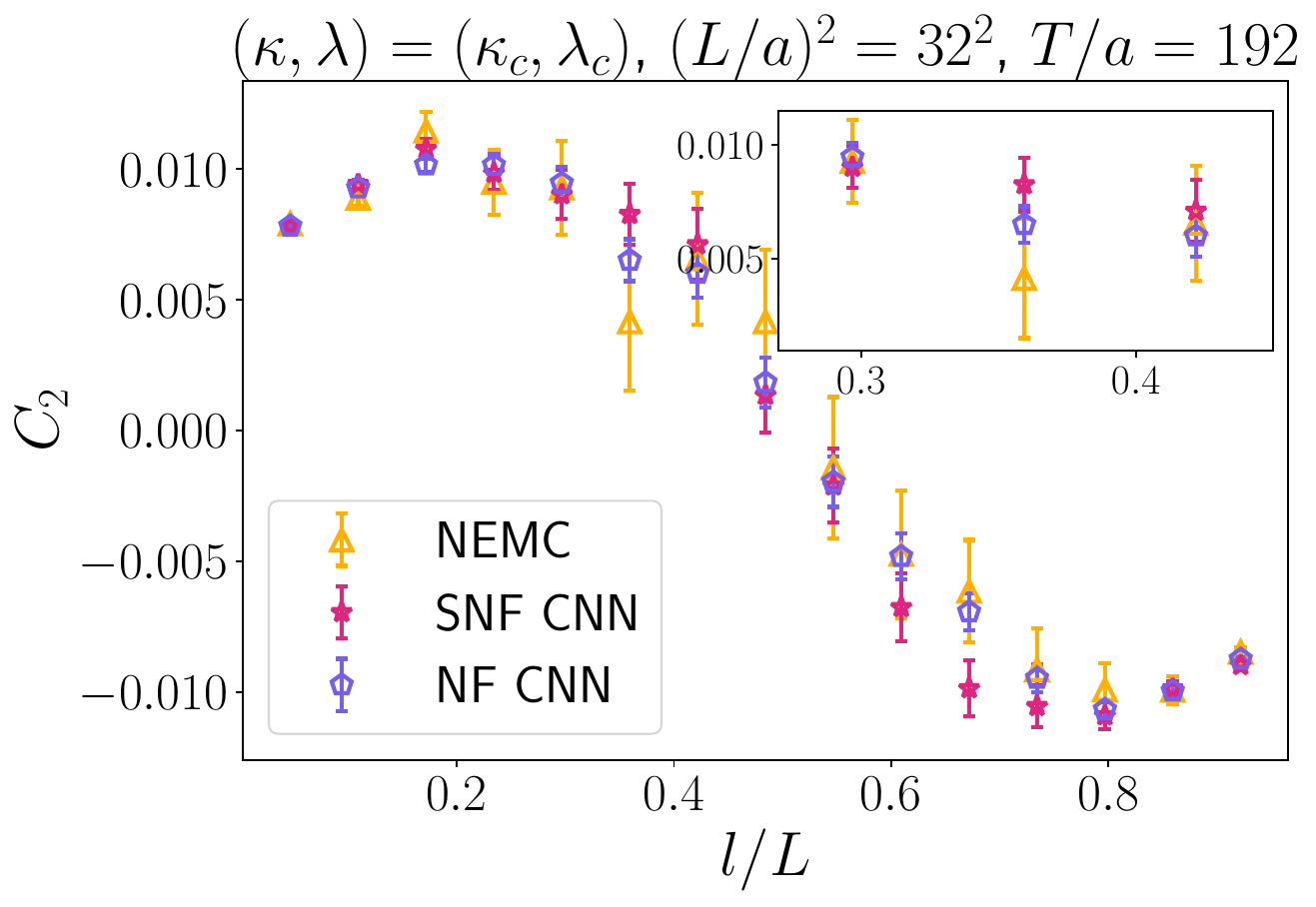}}
\caption{Defect-coupling-layer-based NFs compared with other flow-based samplers. Left panel: total simulation time to reach a given accuracy. Right panel: entropic c-function computed with different samplers at fixed statistics.}
\label{fig:results}
\end{figure}

We compare our proposed approach to other two flow-based samplers, which, differently from NFs, make use of a non-equilibrium Markov chain Monte Carlo process to flow from the prior to the target. Such approaches, named non-equilibrium Monte Carlo (NEMC) and Stochastic Normalizing Flows (SNFs) have well understood scaling properties~\cite{Bonanno:2024udh, Bulgarelli:2024brv, Bonanno:2025pdp}, and have been proven to be suited for large-scale simulations, also for entropic c-function calculations~\cite{Bulgarelli:2024onj}. At the same time, however, Monte Carlo updates introduce a significant computational overhead compared to coupling layers.

Figure~\ref{fig:results}, left panel, shows a study of the computational cost to reach a given accuracy in the determination of the entropic c-function, as a function of the volume of the lattice, in the $(2+1)$-dimensional $\phi^4$ theory at the critical point. In particular, every point represents the integrated cost to perform the simulations up to the volume considered. The number of hours includes training of the models, thermalization and sampling.

In the whole range of volumes explored, the NF consistently outperforms the Monte-Carlo-based architecture, being progressively better for larger volumes. This is the main result of our work, as it proves our approach to be a highly competitive tool for the study of entanglement entropy, and more in general defect-based observables in lattice field theories, for a range of volumes that would be prohibitive for standard NF-based architectures.

In fig.~\ref{fig:results}, left panel, we also show how, for fixed statistics, NFs provide a more accurate estimate of the critical entropic c-function, for all the values of $l$.

\section{Conclusions}

In~\cite{Bulgarelli:2024yrz} we proposed the \textit{defect coupling layer} as an algorithmic tool to study entanglement-related quantities in lattice field theories. The philosophy of this approach closely resembles that of equivariant flows, in which symmetries of the system are built directly into the model rather than learned from data~\cite{Kanwar:2020xzo}. This strategy has been shown to substantially reduce training costs and improve performance~\cite{Kreit:2025cos, Schuh:2025gks, Vega:2025hgz}. In our case, we select \textit{a priori} the relevant portion of the lattice where the model acts. This not only reduces training time but also leads to remarkable transfer-learning capabilities: instead of learning correlations tied to the particular parameter regime in which the model is trained, the NF is found to learn only local geometric features.

It is important to emphasize that the defect coupling layer \textit{per se} does not resolve the scaling problems of NFs with the number of degrees of freedom. Rather, it reduces the computational cost of the simulations to the point where, for a wide range of volumes, NFs outperform traditional methods. For theories with significantly more d.o.f.\ we expect the scaling properties of non-equilibrium approaches to make them more competitive compared to NFs: in such cases, defect coupling layers can be used as a key component of SNFs, which are characterized by well understood scaling properties~\cite{Bulgarelli:2024brv, Bonanno:2025pdp} and can be systematically improved through coupling layers.

Several quantities in lattice field theory require the study of systems with defects. While we foresee potential applications in the study of other entanglement-related quantities, such as excited states entanglement~\cite{Amorosso:2024leg, Amorosso:2024glf} or negativity~\cite{Alba:2013mg, Florio:2023mzk}, just to name a few, our approach is more general and can be applied in different physical setups. An example is the study of topology in $\mathrm{SU}(3)$ lattice gauge theory~\cite{Bonanno:2025pdp}. We also envision applications to the study of interfaces in spin models as well as gauge theories~\cite{Caselle:2007yc}.

\section*{Acknowledgments}
\noindent
This work is funded by the European Union’s Horizon Europe Framework Programme (HORIZON) under the ERA Chair scheme with grant agreement no.\ 101087126.
    This work is supported with funds from the Ministry of Science, Research and Culture of the State of Brandenburg within the Center for Quantum Technologies and Applications (CQTA).
    This project was funded by the Deutsche Forschungsgemeinschaft (DFG, German Research Foundation) as part of the CRC 1639 NuMeriQS – project no. 511713970 and under Germany’s Excellence Strategy – Cluster of Excellence Matter and Light for Quantum Computing (ML4Q) EXC 2004/1 – 390534769.
    A.~N. and E.~C. acknowledge support and A.~B. acknowledges partial support by the Simons Foundation grant 994300 (Simons Collaboration on Confinement and QCD Strings).  A.~N. acknowledges support from the European Union -- Next Generation EU, Mission 4 Component 1, CUP D53D23002970006, under the Italian PRIN ``Progetti di Ricerca di Rilevante Interesse Nazionale – Bando 2022'' prot. 2022ZTPK4E. A.~B., E.~C., A.~N. and M.~P. acknowledge support from the SFT Scientific Initiative of INFN. K.~A.~N. is supported by the Deutsche Forschungsgemeinschaft (DFG, German Research Foundation) as part of the CRC 1639 NuMeriQS – project no. 511713970.
    S.~N. is supported by the German Ministry for Education and Research (BMBF) as BIFOLD – Berlin Institute for the Foundations of Learning and Data (BIFOLD24B),
    and by the European Union’s HORIZON MSCA Doctoral Networks programme project AQTIVATE (101072344).
    The numerical simulations were run on machines of the Consorzio Interuniversitario per il Calcolo Automatico dell'Italia Nord Orientale (CINECA).

\bibliographystyle{JHEPjus}
\bibliography{pos2025}

\providecommand{\href}[2]{#2}\begingroup\begin{thebibliography}{10}

\bibitem{Bulgarelli:2024yrz}
A.~Bulgarelli, E.~Cellini, K.~Jansen, S.~K\"uhn, A.~Nada, S.~Nakajima et~al.,
  \emph{{Flow-Based Sampling for Entanglement Entropy and the Machine Learning
  of Defects}},
  \href{http://dx.doi.org/10.1103/PhysRevLett.134.151601}{\emph{Phys. Rev.
  Lett.} {\bfseries 134} (2025) 151601},
  [\href{https://arxiv.org/abs/2410.14466}{{\ttfamily 2410.14466}}].

\bibitem{Calabrese:2004eu}
P.~Calabrese and J.~L. Cardy, \emph{{Entanglement entropy and quantum field
  theory}}, \href{http://dx.doi.org/10.1088/1742-5468/2004/06/P06002}{\emph{J.
  Stat. Mech.} {\bfseries 0406} (2004) P06002},
  [\href{https://arxiv.org/abs/hep-th/0405152}{{\ttfamily hep-th/0405152}}].

\bibitem{Vidal:2002rm}
G.~Vidal, J.~I. Latorre, E.~Rico and A.~Kitaev, \emph{{Entanglement in quantum
  critical phenomena}},
  \href{http://dx.doi.org/10.1103/PhysRevLett.90.227902}{\emph{Phys. Rev.
  Lett.} {\bfseries 90} (2003) 227902},
  [\href{https://arxiv.org/abs/quant-ph/0211074}{{\ttfamily
  quant-ph/0211074}}].

\bibitem{Ryu:2006bv}
S.~Ryu and T.~Takayanagi, \emph{{Holographic derivation of entanglement entropy
  from AdS/CFT}},
  \href{http://dx.doi.org/10.1103/PhysRevLett.96.181602}{\emph{Phys. Rev.
  Lett.} {\bfseries 96} (2006) 181602},
  [\href{https://arxiv.org/abs/hep-th/0603001}{{\ttfamily hep-th/0603001}}].

\bibitem{Klebanov:2007ws}
I.~R. Klebanov, D.~Kutasov and A.~Murugan, \emph{{Entanglement as a probe of
  confinement}},
  \href{http://dx.doi.org/10.1016/j.nuclphysb.2007.12.017}{\emph{Nucl. Phys. B}
  {\bfseries 796} (2008) 274},
  [\href{https://arxiv.org/abs/0709.2140}{{\ttfamily 0709.2140}}].

\bibitem{Buividovich:2008gq}
P.~V. Buividovich and M.~I. Polikarpov, \emph{{Entanglement entropy in gauge
  theories and the holographic principle for electric strings}},
  \href{http://dx.doi.org/10.1016/j.physletb.2008.10.032}{\emph{Phys. Lett. B}
  {\bfseries 670} (2008) 141},
  [\href{https://arxiv.org/abs/0806.3376}{{\ttfamily 0806.3376}}].

\bibitem{Casini:2013rba}
H.~Casini, M.~Huerta and J.~A. Rosabal, \emph{{Remarks on entanglement entropy
  for gauge fields}},
  \href{http://dx.doi.org/10.1103/PhysRevD.89.085012}{\emph{Phys. Rev. D}
  {\bfseries 89} (2014) 085012},
  [\href{https://arxiv.org/abs/1312.1183}{{\ttfamily 1312.1183}}].

\bibitem{Ghosh:2015iwa}
S.~Ghosh, R.~M. Soni and S.~P. Trivedi, \emph{{On The Entanglement Entropy For
  Gauge Theories}},
  \href{http://dx.doi.org/10.1007/JHEP09(2015)069}{\emph{JHEP} {\bfseries 09}
  (2015) 069}, [\href{https://arxiv.org/abs/1501.02593}{{\ttfamily
  1501.02593}}].

\bibitem{Albergo:2019eim}
M.~S. Albergo, G.~Kanwar and P.~E. Shanahan, \emph{{Flow-based generative
  models for Markov chain Monte Carlo in lattice field theory}},
  \href{http://dx.doi.org/10.1103/PhysRevD.100.034515}{\emph{Phys. Rev. D}
  {\bfseries 100} (2019) 034515},
  [\href{https://arxiv.org/abs/1904.12072}{{\ttfamily 1904.12072}}].

\bibitem{Nicoli:2019gun}
K.~A. Nicoli, S.~Nakajima, N.~Strodthoff, W.~Samek, K.-R. M{\"u}ller and
  P.~Kessel, \emph{{Asymptotically unbiased estimation of physical observables
  with neural samplers}},
  \href{http://dx.doi.org/10.1103/PhysRevE.101.023304}{\emph{Phys. Rev. E}
  {\bfseries 101} (2020) 023304},
  [\href{https://arxiv.org/abs/1910.13496}{{\ttfamily 1910.13496}}].

\bibitem{Kanwar:2020xzo}
G.~Kanwar, M.~S. Albergo, D.~Boyda, K.~Cranmer, D.~C. Hackett, S.~Racani{\`e}re
  et~al., \emph{{Equivariant flow-based sampling for lattice gauge theory}},
  \href{http://dx.doi.org/10.1103/PhysRevLett.125.121601}{\emph{Phys. Rev.
  Lett.} {\bfseries 125} (2020) 121601},
  [\href{https://arxiv.org/abs/2003.06413}{{\ttfamily 2003.06413}}].

\bibitem{Bacchio:2022vje}
S.~Bacchio, P.~Kessel, S.~Schaefer and L.~Vaitl, \emph{{Learning trivializing
  gradient flows for lattice gauge theories}},
  \href{http://dx.doi.org/10.1103/PhysRevD.107.L051504}{\emph{Phys. Rev. D}
  {\bfseries 107} (2023) L051504},
  [\href{https://arxiv.org/abs/2212.08469}{{\ttfamily 2212.08469}}].

\bibitem{Cranmer:2023xbe}
K.~Cranmer, G.~Kanwar, S.~Racani{\`e}re, D.~J. Rezende and P.~E. Shanahan,
  \emph{{Advances in machine-learning-based sampling motivated by lattice
  quantum chromodynamics}},
  \href{http://dx.doi.org/10.1038/s42254-023-00616-w}{\emph{Nature Rev. Phys.}
  {\bfseries 5} (2023) 526},
  [\href{https://arxiv.org/abs/2309.01156}{{\ttfamily 2309.01156}}].

\bibitem{Caselle:2023mvh}
M.~Caselle, E.~Cellini and A.~Nada, \emph{{Sampling the lattice Nambu-Goto
  string using Continuous Normalizing Flows}},
  \href{http://dx.doi.org/10.1007/JHEP02(2024)048}{\emph{JHEP} {\bfseries 02}
  (2024) 048}, [\href{https://arxiv.org/abs/2307.01107}{{\ttfamily
  2307.01107}}].

\bibitem{Albandea:2023wgd}
D.~Albandea, L.~Del~Debbio, P.~Hern{\'a}ndez, R.~Kenway, J.~Marsh~Rossney and
  A.~Ramos, \emph{{Learning trivializing flows}},
  \href{http://dx.doi.org/10.1140/epjc/s10052-023-11838-8}{\emph{Eur. Phys. J.
  C} {\bfseries 83} (2023) 676},
  [\href{https://arxiv.org/abs/2302.08408}{{\ttfamily 2302.08408}}].

\bibitem{Caselle:2024ent}
M.~Caselle, E.~Cellini and A.~Nada, \emph{{Numerical determination of the width
  and shape of the effective string using Stochastic Normalizing Flows}},
  \href{http://dx.doi.org/10.1007/JHEP02(2025)090}{\emph{JHEP} {\bfseries 02}
  (2025) 090}, [\href{https://arxiv.org/abs/2409.15937}{{\ttfamily
  2409.15937}}].

\bibitem{Singha:2025lsd}
A.~Singha, E.~Cellini, K.~A. Nicoli, K.~Jansen, S.~K{\"u}hn and S.~Nakajima,
  \emph{{Multilevel Generative Samplers for Investigating Critical Phenomena}},
   in \emph{{International Conference on Learning Representations}}, 3, 2025,
  \href{https://arxiv.org/abs/2503.08918}{{\ttfamily 2503.08918}}.

\bibitem{Kreit:2025cos}
J.~Kreit, D.~Schuh, K.~A. Nicoli and L.~Funcke, \emph{{SESaMo:
  Symmetry-Enforcing Stochastic Modulation for Normalizing Flows}},
  \href{https://arxiv.org/abs/2505.19619}{{\ttfamily 2505.19619}}.

\bibitem{Schuh:2025gks}
D.~Schuh, J.~Kreit, E.~Berkowitz, L.~Funcke, T.~Luu, K.~A. Nicoli et~al.,
  \emph{{Simulating Correlated Electrons with Symmetry-Enforced Normalizing
  Flows}},  \href{https://arxiv.org/abs/2506.17015}{{\ttfamily 2506.17015}}.

\bibitem{Vega:2025hgz}
O.~Vega, J.~Komijani, A.~El-Khadra and M.~Marinkovic, \emph{{Group-Equivariant
  Diffusion Models for Lattice Field Theory}},
  \href{https://arxiv.org/abs/2510.26081}{{\ttfamily 2510.26081}}.

\bibitem{Caselle:2016wsw}
M.~Caselle, G.~Costagliola, A.~Nada, M.~Panero and A.~Toniato,
  \emph{{Jarzynski{\textquoteright}s theorem for lattice gauge theory}},
  \href{http://dx.doi.org/10.1103/PhysRevD.94.034503}{\emph{Phys. Rev. D}
  {\bfseries 94} (2016) 034503},
  [\href{https://arxiv.org/abs/1604.05544}{{\ttfamily 1604.05544}}].

\bibitem{Alba:2016bcp}
V.~Alba, \emph{{Out-of-equilibrium protocol for R{\'e}nyi entropies via the
  Jarzynski equality}},
  \href{http://dx.doi.org/10.1103/PhysRevE.95.062132}{\emph{Phys. Rev. E}
  {\bfseries 95} (2017) 062132},
  [\href{https://arxiv.org/abs/1609.02157}{{\ttfamily 1609.02157}}].

\bibitem{Caselle:2018kap}
M.~Caselle, A.~Nada and M.~Panero, \emph{{QCD thermodynamics from lattice
  calculations with nonequilibrium methods: The SU(3) equation of state}},
  \href{http://dx.doi.org/10.1103/PhysRevD.98.054513}{\emph{Phys. Rev. D}
  {\bfseries 98} (2018) 054513},
  [\href{https://arxiv.org/abs/1801.03110}{{\ttfamily 1801.03110}}].

\bibitem{Nicoli:2020njz}
K.~A. Nicoli, C.~J. Anders, L.~Funcke, T.~Hartung, K.~Jansen, P.~Kessel et~al.,
  \emph{{Estimation of Thermodynamic Observables in Lattice Field Theories with
  Deep Generative Models}},
  \href{http://dx.doi.org/10.1103/PhysRevLett.126.032001}{\emph{Phys. Rev.
  Lett.} {\bfseries 126} (2021) 032001},
  [\href{https://arxiv.org/abs/2007.07115}{{\ttfamily 2007.07115}}].

\bibitem{Bulgarelli:2023ofi}
A.~Bulgarelli and M.~Panero, \emph{{Entanglement entropy from non-equilibrium
  Monte Carlo simulations}},
  \href{http://dx.doi.org/10.1007/JHEP06(2023)030}{\emph{JHEP} {\bfseries 06}
  (2023) 030}, [\href{https://arxiv.org/abs/2304.03311}{{\ttfamily
  2304.03311}}].

\bibitem{Bulgarelli:2024onj}
A.~Bulgarelli and M.~Panero, \emph{{Duality transformations and the
  entanglement entropy of gauge theories}},
  \href{http://dx.doi.org/10.1007/JHEP06(2024)041}{\emph{JHEP} {\bfseries 06}
  (2024) 041}, [\href{https://arxiv.org/abs/2404.01987}{{\ttfamily
  2404.01987}}].

\bibitem{Bialas:2024gha}
P.~Bia{\l}as, P.~Korcyl, T.~Stebel and D.~Zapolski, \emph{{R{\'e}nyi
  entanglement entropy of a spin chain with generative neural networks}},
  \href{http://dx.doi.org/10.1103/PhysRevE.110.044116}{\emph{Phys. Rev. E}
  {\bfseries 110} (2024) 044116},
  [\href{https://arxiv.org/abs/2406.06193}{{\ttfamily 2406.06193}}].

\bibitem{Bulgarelli:2025riv}
A.~Bulgarelli, M.~Caselle, A.~Nada and M.~Panero, \emph{{Casimir effect in
  critical $\mathrm{O}(N)$ models from non-equilibrium Monte Carlo
  simulations}},  \href{https://arxiv.org/abs/2505.20403}{{\ttfamily
  2505.20403}}.

\bibitem{Caselle:2022acb}
M.~Caselle, E.~Cellini, A.~Nada and M.~Panero, \emph{{Stochastic normalizing
  flows as non-equilibrium transformations}},
  \href{http://dx.doi.org/10.1007/JHEP07(2022)015}{\emph{JHEP} {\bfseries 07}
  (2022) 015}, [\href{https://arxiv.org/abs/2201.08862}{{\ttfamily
  2201.08862}}].

\bibitem{Bonanno:2024udh}
C.~Bonanno, A.~Nada and D.~Vadacchino, \emph{{Mitigating topological freezing
  using out-of-equilibrium simulations}},
  \href{http://dx.doi.org/10.1007/JHEP04(2024)126}{\emph{JHEP} {\bfseries 04}
  (2024) 126}, [\href{https://arxiv.org/abs/2402.06561}{{\ttfamily
  2402.06561}}].

\bibitem{Bulgarelli:2024brv}
A.~Bulgarelli, E.~Cellini and A.~Nada, \emph{{Scaling of stochastic normalizing
  flows in SU(3) lattice gauge theory}},
  \href{http://dx.doi.org/10.1103/PhysRevD.111.074517}{\emph{Phys. Rev. D}
  {\bfseries 111} (2025) 074517},
  [\href{https://arxiv.org/abs/2412.00200}{{\ttfamily 2412.00200}}].

\bibitem{Bonanno:2025pdp}
C.~Bonanno, A.~Bulgarelli, E.~Cellini, A.~Nada, D.~Panfalone, D.~Vadacchino
  et~al., \emph{{Scaling flow-based approaches for topology sampling in
  $\mathrm{SU}(3)$ gauge theory}},
  \href{https://arxiv.org/abs/2510.25704}{{\ttfamily 2510.25704}}.

\bibitem{Rezende:2015}
D.~Rezende and S.~Mohamed, \emph{Variational inference with normalizing flows},
   in \emph{International conference on machine learning}, pp.~1530--1538,
  PMLR, 2015.

\bibitem{Casini:2009sr}
H.~Casini and M.~Huerta, \emph{{Entanglement entropy in free quantum field
  theory}}, \href{http://dx.doi.org/10.1088/1751-8113/42/50/504007}{\emph{J.
  Phys. A} {\bfseries 42} (2009) 504007},
  [\href{https://arxiv.org/abs/0905.2562}{{\ttfamily 0905.2562}}].

\bibitem{Buividovich:2008kq}
P.~V. Buividovich and M.~I. Polikarpov, \emph{{Numerical study of entanglement
  entropy in SU(2) lattice gauge theory}},
  \href{http://dx.doi.org/10.1016/j.nuclphysb.2008.04.024}{\emph{Nucl. Phys. B}
  {\bfseries 802} (2008) 458},
  [\href{https://arxiv.org/abs/0802.4247}{{\ttfamily 0802.4247}}].

\bibitem{Itou:2015cyu}
E.~Itou, K.~Nagata, Y.~Nakagawa, A.~Nakamura and V.~I. Zakharov,
  \emph{{Entanglement in Four-Dimensional SU(3) Gauge Theory}},
  \href{http://dx.doi.org/10.1093/ptep/ptw050}{\emph{PTEP} {\bfseries 2016}
  (2016) 061B01}, [\href{https://arxiv.org/abs/1512.01334}{{\ttfamily
  1512.01334}}].

\bibitem{Rabenstein:2018bri}
A.~Rabenstein, N.~Bodendorfer, P.~Buividovich and A.~Sch\"afer, \emph{{Lattice
  study of R\'enyi entanglement entropy in $SU(N_c)$ lattice Yang-Mills theory
  with $N_c = 2, 3, 4$}},
  \href{http://dx.doi.org/10.1103/PhysRevD.100.034504}{\emph{Phys. Rev. D}
  {\bfseries 100} (2019) 034504},
  [\href{https://arxiv.org/abs/1812.04279}{{\ttfamily 1812.04279}}].

\bibitem{DelDebbio:2021qwf}
L.~Del~Debbio, J.~Marsh~Rossney and M.~Wilson, \emph{{Efficient modeling of
  trivializing maps for lattice \ensuremath{\phi^4} theory using normalizing
  flows: A first look at scalability}},
  \href{http://dx.doi.org/10.1103/PhysRevD.104.094507}{\emph{Phys. Rev. D}
  {\bfseries 104} (2021) 094507},
  [\href{https://arxiv.org/abs/2105.12481}{{\ttfamily 2105.12481}}].

\bibitem{Abbott:2022zsh}
R.~Abbott et~al., \emph{{Aspects of scaling and scalability for flow-based
  sampling of lattice QCD}},
  \href{http://dx.doi.org/10.1140/epja/s10050-023-01154-w}{\emph{Eur. Phys. J.
  A} {\bfseries 59} (2023) 257},
  [\href{https://arxiv.org/abs/2211.07541}{{\ttfamily 2211.07541}}].

\bibitem{Abbott:2023thq}
R.~Abbott et~al., \emph{{Normalizing flows for lattice gauge theory in
  arbitrary space-time dimension}},
  \href{https://arxiv.org/abs/2305.02402}{{\ttfamily 2305.02402}}.

\bibitem{Bianchi:2015liz}
L.~Bianchi, M.~Meineri, R.~C. Myers and M.~Smolkin, \emph{{R\'enyi entropy and
  conformal defects}},
  \href{http://dx.doi.org/10.1007/JHEP07(2016)076}{\emph{JHEP} {\bfseries 07}
  (2016) 076}, [\href{https://arxiv.org/abs/1511.06713}{{\ttfamily
  1511.06713}}].

\bibitem{Dinh:2017}
L.~Dinh, J.~Sohl-Dickstein and S.~Bengio, \emph{Density estimation using {R}eal
  {NVP}},  in \emph{International Conference on Learning Representations},
  2017, \href{https://arxiv.org/abs/1605.08803}{{\ttfamily 1605.08803}}.

\bibitem{Amorosso:2024leg}
R.~Amorosso, S.~Syritsyn and R.~Venugopalan, \emph{{Entanglement entropy of a
  color flux tube in (2+1)D Yang-Mills theory}},
  \href{http://dx.doi.org/10.1007/JHEP12(2024)177}{\emph{JHEP} {\bfseries 12}
  (2024) 177}, [\href{https://arxiv.org/abs/2410.00112}{{\ttfamily
  2410.00112}}].

\bibitem{Amorosso:2024glf}
R.~Amorosso, S.~Syritsyn and R.~Venugopalan, \emph{{Entanglement entropy of a
  color flux tube in (1+1)D Yang{\textendash}Mills theory}},
  \href{http://dx.doi.org/10.1016/j.physletb.2025.139806}{\emph{Phys. Lett. B}
  {\bfseries 868} (2025) 139806},
  [\href{https://arxiv.org/abs/2411.12818}{{\ttfamily 2411.12818}}].

\bibitem{Alba:2013mg}
V.~Alba, \emph{{Entanglement negativity and conformal field theory: a Monte
  Carlo study}},
  \href{http://dx.doi.org/10.1088/1742-5468/2013/05/P05013}{\emph{J. Stat.
  Mech.} {\bfseries 1305} (2013) P05013},
  [\href{https://arxiv.org/abs/1302.1110}{{\ttfamily 1302.1110}}].

\bibitem{Florio:2023mzk}
A.~Florio, \emph{{Two-fermion negativity and confinement in the Schwinger
  model}}, \href{http://dx.doi.org/10.1103/PhysRevD.109.L071501}{\emph{Phys.
  Rev. D} {\bfseries 109} (2024) L071501},
  [\href{https://arxiv.org/abs/2312.05298}{{\ttfamily 2312.05298}}].

\bibitem{Caselle:2007yc}
M.~Caselle, M.~Hasenbusch and M.~Panero, \emph{{The Interface free energy:
  Comparison of accurate Monte Carlo results for the 3D Ising model with
  effective interface models}},
  \href{http://dx.doi.org/10.1088/1126-6708/2007/09/117}{\emph{JHEP} {\bfseries
  09} (2007) 117}, [\href{https://arxiv.org/abs/0707.0055}{{\ttfamily
  0707.0055}}].

\end{thebibliography}\endgroup

\end{document}